\documentclass[twocolumn,aps,prl,floats,showpacs]{revtex4}
\usepackage{epsfig}
\usepackage{dcolumn}
\usepackage{bm}
\begin{document}

\title{Aging Effects Across the Metal-Insulator Transition in Two Dimensions}
\author{J. Jaroszy\'nski}
\affiliation{National High Magnetic Field Laboratory, Florida State
University, Tallahassee, Florida 32310, USA}
\author{Dragana  Popovi\'c}
\email{dragana@magnet.fsu.edu} \affiliation{National High Magnetic Field Laboratory, Florida State University,
Tallahassee, Florida 32310, USA}
\date{\today}

\begin{abstract}
Aging effects in the relaxations of conductivity of a
two-dimensional electron system in Si have been studied as a
function of carrier density.  They reveal an abrupt change in the
nature of the glassy phase at the metal-insulator transition
(MIT): (a) while full aging is observed in the insulating regime,
there are significant departures from full aging on the metallic
side of the MIT, before the glassy phase disappears completely at
a higher density $n_g$; (b) the amplitude of the relaxations peaks
just below the MIT, and it is strongly suppressed in the
insulating phase.  Other aspects of aging, including large
non-Gaussian noise and similarities to spin glasses, also have been
discussed.
\end{abstract}


\maketitle

The glassy freezing of electrons, resulting from the competition
between disorder and strong electron-electron interactions, may be
crucial for our understanding of the behavior of many materials
near the metal-insulator transition
(MIT)~\cite{Mir-Dob-review}.  The dynamics of such electron or
Coulomb glasses~\cite{eglass}, however, remains poorly understood,
and experimental studies~\cite{earlyg,films-Zvi,films3} are still
relatively scarce.  In two dimensions (2D), moreover, even the
very existence of the metal
and the MIT have been questioned.  Although there is
now considerable experimental evidence in favor of such a
transition~\cite{2DMIT-reviews}, there is still no general
agreement even on the fundamental driving force for the 2D MIT.
Theoretical proposals describing the 2D MIT as the melting of a
Coulomb glass~\cite{MIT-glassothers,Vlad-MITglass,Darko} have
found support in
observations and studies of
glassiness~\cite{SBPRL,JJPRL,relax-PRL,tw-PRL} in a 2D electron
system (2DES) in Si.  Aging~\cite{Struik,aging}, one of the key
characteristics of glassy materials, also has been demonstrated in
this system~\cite{tw-PRL}, but its properties have not been
investigated in detail.  Surprisingly, although consistent with
theory~\cite{Darko}, glassiness sets in on the metallic side of
the MIT~\cite{SBPRL,JJPRL,relax-PRL}, \textit{i.e.} at an electron
density $n_g>n_c$ ($n_c$-- the critical density for the MIT), thus
giving rise to an intermediate, poorly metallic and glassy phase
between the metal and the (glassy) insulator.  Here we focus on
aging effects, which have been instrumental as a probe of complex
nonequilibrium dynamics in many types of materials.  We find an
abrupt change in the aging properties that occurs at $n_c$.  The
observed change in the dynamics at the 2D MIT itself, in addition
to that at $n_g$~\cite{SBPRL,JJPRL,relax-PRL}, might help to
determine the validity of different theoretical pictures.

The system is said to exhibit aging if its response to an external
excitation depends on the system history in addition to the time
$t$.  In electron glasses, aging has been studied mostly and most
easily~\cite{films-Zvi} by looking at the relaxations of
conductivity $\sigma(t)$ towards its equilibrium value $\sigma_0$
after a temporary change of electron density $n_s$ during the
waiting time $t_w$.  Aging is observed if $t_w\ll\tau_{eq}$
($\tau_{eq}$ -- equilibration time), \textit{i.e.} if the system
is not able to reach equilibrium under the new conditions during
$t_w$.  It is manifested in the dependence of $\sigma(t)$ on $t_w$
such that, in those strongly localized systems, the aging function
$\sigma(t,t_w)$ is just a function of $t/t_w$~\cite{films-Zvi}.
This is known as simple, or full aging.  It is interesting that,
in spin glasses, full aging has been demonstrated only relatively
recently~\cite{sg-fullaging}.  In general, however, the existence
of a characteristic time scale $t_w$ does not necessarily imply
simple $t/t_w$ scaling~\cite{LesHouches-2002}.  In the mean-field
models, in fact, two different cases are distinguished:
one, where full aging is expected, and the other, where no $t/t_w$
scaling is expected~\cite{mean-field}. Experimentally, departures
from full aging are common~\cite{Struik,aging}.  In a 2DES in Si,
where $\tau_{eq}\rightarrow\infty$ as temperature $T\rightarrow 0$
(hence glass transition $T_g=0$)~\cite{relax-PRL}, aging was also
observed for $t_w\ll\tau_{eq}(T)$~\cite{tw-PRL} using the
experimental protocol described above.  The goal of this study is
to investigate that aging regime in detail.  In particular, unlike
previous work~\cite{tw-PRL}, here $T$ is kept fixed at 1~K such
that $\tau_{eq}$ is astronomical~\cite{tauhigh} and the 2DES is
always deep in the $t_w\ll\tau_{eq}$ limit; $\sigma(t,t_w)$ are
then explored systematically both as a function of final $n_s$ and
of the difference in densities during and after $t_w$.  Our main
results include: a) $\sigma(t)$ obey a power-law dependence for
$t\lesssim t_w$ and a slower relaxation law for $t\gtrsim t_w$,
thus showing a memory of the time (``age'') $t_w$; b)
$\sigma(t,t_w)$ exhibit full aging for $n_s<n_c$; c) as $n_s$
increases above $n_c$, there is an increasingly strong departure
from full aging that reaches maximum at $n_s\simeq n_g$; d) for a
given $t_w$, the amplitude of $\sigma(t)/\sigma_0$ has a peak at
$n_s\lesssim n_c$, reflecting an interesting and surprising
\textit{suppression} of the relaxations on the insulating side of
the 2D MIT.

The experiment was performed in a $^{3}$He system (base $T=0.24$~K) on the same (100)-Si metal-oxide-semiconductor field-effect transistors that were used in previous studies~\cite{relax-PRL,tw-PRL}.  The two devices (A and B, with $1\times 90~\mu$m$^2$ and $2\times 50~\mu$m$^2$ length$\,\times\,$width, respectively) had a 50~nm oxide thickness, and a relatively large amount of disorder (the 4.2~K peak mobility $\approx$~0.06~m$^2$/Vs with the substrate (back-gate) bias~\cite{AFS} $V_{sub}=-2$~V).  $n_s$ was varied by the gate voltage $V_g$, such that
$n_s(10^{11}$cm$^{-2})=4.31(V_g[$V$]-6.3)$;
$n_g(10^{11}$cm$^{-2})=(7.5\pm 0.3)$ and
$n_c(10^{11}$cm$^{-2})=(4.5\pm 0.4)$, where $n_g$ was determined from the onset of slow, correlated dynamics in noise and $n_c$ from $\sigma(n_s,T)$ measurements on both metallic and insulating sides~\cite{SBPRL,JJPRL}. The devices
and the standard ac lock-in technique (typically 13 Hz; 5--10~$\mu$V excitation voltage)
were described in more detail
elsewhere~\cite{SBPRL}.  The two samples exhibited an almost
identical behavior.  Unless noted otherwise, the data presented
below were obtained on sample A.

\begin{figure}
\centerline{\epsfig{file=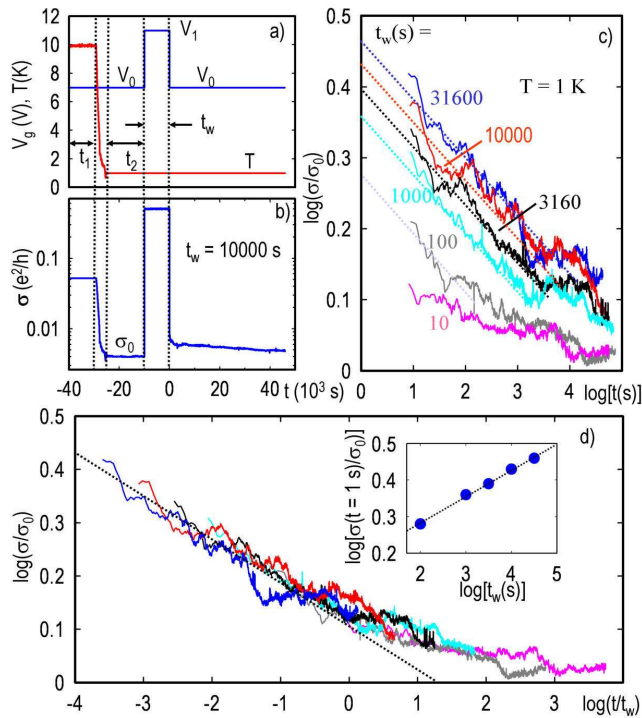,width=8.5cm,clip=}}
\caption{(color online)
(a) $V_g(t)$ and $T(t)$ in a typical experimental protocol, which
always starts with the 2DES in equilibrium at 10~K~\cite{tw-PRL}.
The results do not depend on the cooling time (varied from 30
minutes to 10 hours), nor on $t_1$ and $t_2$ (varied from 5
minutes to 8 hours each). (b) The corresponding $\sigma(t)$.  The
relaxation of $\sigma$ during $t_w$ is too small to be seen on
this scale.  (c)  $\sigma(t>0)$ for several $t_w$, as shown;
$V_{0}=7.0$~V [$n_{0}(10^{11}$cm$^{-2})=3.02<n_c$], $V_{1}=11$~V
[$n_1(10^{11}$cm$^{-2})=20.26$].  The data are not shown for the
first few seconds (comparable to our sampling time $\sim 1$~s). The
dotted lines are linear fits for $t\leq t_w$.  (d) The same data
as in (c) but plotted \textit{vs.} $t/t_w$.  The dotted line is a
fit for $t\leq t_w$ with the slope $-\alpha=-0.081\pm 0.005$.
Inset: $\sigma(t=1$s$)/\sigma_0$ \textit{vs.} $t_w$.  The dotted
line is a fit with the slope $\alpha=0.076\pm 0.005$.
 \label{fig:full}}
\end{figure}

We employ the so-called ``gate protocol''~\cite{films-Zvi}, where $V_g$ is changed from an initial value $V_0$ (density $n_0$), where the 2DES is in equilibrium, to another one, $V_1$ (density $n_1$), where the system attempts to equilibrate during $t_w$, but $t_w\ll\tau_{eq}$ [Figs.~\ref{fig:full}(a) and \ref{fig:full}(b)].  After $V_g$ is changed back to $V_0$ at $t=0$,
%
%
the relaxation of the ``excess'' conductivity $\sigma(V_0,t)/\sigma(V_0)$ is studied.  Figure~\ref{fig:full}(c) shows some $\sigma(t)$ measured for several $t_w$ and $n_0<n_c$.
Although $n_1>n_g$,
small amplitude relaxation during $t_w$ is still visible~\cite{relax-PRL,tw-PRL,kfl}.  It is clear that $t_w$ has a significant effect on $\sigma(t)$.  In fact, all the $\sigma(t,t_w)$ data can be collapsed onto a single curve simply by rescaling the time axis by $t_w$ [Fig.~\ref{fig:full}(d)].  Therefore, in this case, the system exhibits full aging at least up to $t\approx(2$--$3)t_w$.  We note that the relaxations can be described by a power law $\sigma(t)/\sigma_0\propto (t/t_w)^{-\alpha}$ for times up to about $t_w$, followed by a slower relaxation at longer $t$.  This means that the memory of $t_w$ is imprinted on the form of each $\sigma(t)$.  As a consistency check, the fits to the individual $\sigma(t)$ for $t\leq t_w$ [see Fig.~\ref{fig:full}(c)] yield the initial amplitudes of the relaxation $\sigma(t=1$~s$)/\sigma_0\propto t_{w}^{\alpha}$~\cite{tw-comment} with the same value of $\alpha$ (Fig.~\ref{fig:full}(d) inset).

\begin{figure}
\centerline{\epsfig{file=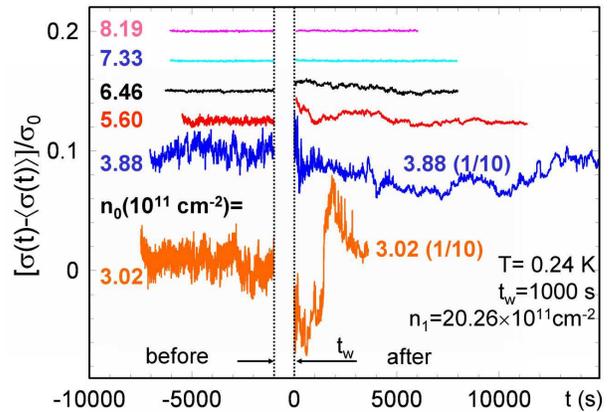,width=8.0cm,clip=}}
\caption{(color online)
Noise in $\sigma$ for
several $n_0$,
before and after a change of $n_s$ to
$n_1$ during $t_{w}$.  The slowly relaxing
background $<\sigma(t)>$ has been subtracted from the upper 3 ``after''
curves. The data are shifted for clarity.  For the two lowest $n_0$, the
signals measured after $t_w$ were divided by 10.
 \label{fig:noise}}
\end{figure}

The quality of the data collapse is difficult to estimate with
high precision since $\sigma(t)$ are quite noisy. We argue that
the large non-Gaussian noise seen in Fig.~\ref{fig:full}(c) is
intrinsic to aging, similar to other materials out of
equilibrium~\cite{LesHouches-2002,Zvi-nonlinear}.  Indeed, Fig.~\ref{fig:noise}
shows that, for $n_s<n_g$,
the noise after a temporary $n_s$ change is much larger than the
%
%
noise before the change.  At the lowest $n_0$, this difference
amounts to more than an order of magnitude.  While a more detailed
comparison of the ``before'' and ``after'' noise will be presented
elsewhere, we note
the following. First, the
noise in the aging regime grows with decreasing $T$, while at the
same time the relaxations $\sigma(t)/\sigma_0$ become smaller.
Therefore, in order to optimize the signal to intrinsic sample noise for the study
of aging, most of the data
were taken
at 1~K.  Second, they
show that the previous noise
studies in a 2DES~\cite{SBPRL,JJPRL} were actually performed in
the aging regime, since $V_g$ was varied at $T\sim 1$~K albeit in
small steps ($\Delta V_g=0.1$ and 0.01~V, or smaller, in
Refs.~\cite{SBPRL} and \cite{JJPRL}, respectively). Such small
$\Delta V_g$ did not produce any visible relaxations within the
noise, consistent with the results below (Fig.~\ref{fig:DeltaV}).

\begin{figure}
\centerline{\epsfig{file=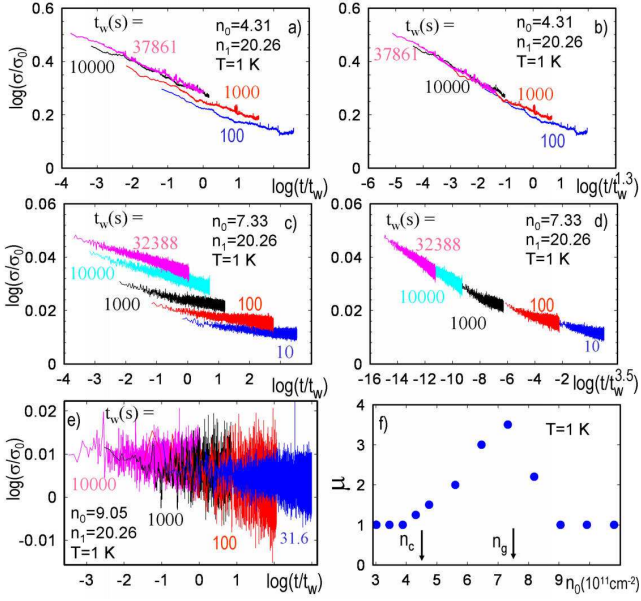,width=8.5cm,clip=}}
\caption{(color online) (a), (c), (e) Relaxations for different
$n_0 (10^{11}$cm$^{-2})$ and a fixed
$n_1 (10^{11}$cm$^{-2})$,
scaled with the waiting time $t_w$.  (b), (d) Scaling with
$t_{w}^{\mu}$ improves the collapse of the
data.  (f) $\mu$
\textit{vs.} $n_0$.  $\mu$ does not depend on $n_1$.\label{fig:mu}}
\end{figure}

\begin{figure}
\centerline{\epsfig{file=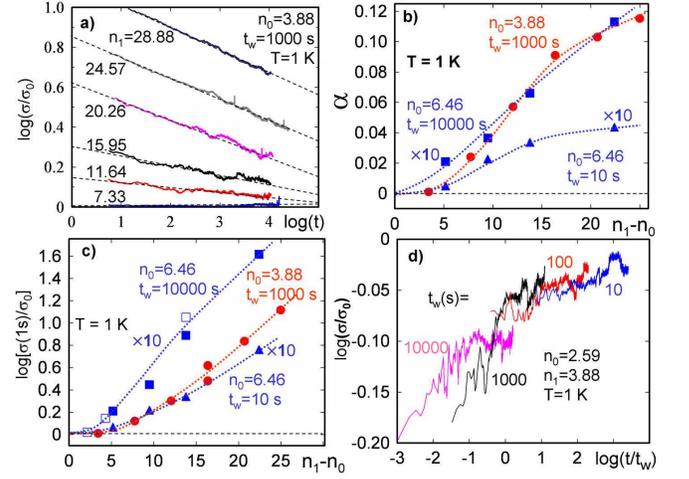,width=8.5cm,clip=}}
\caption{(color online) (a) $\sigma(t)$ for
fixed $n_0 (10^{11}$cm$^{-2})$,
$t_w$, and several $n_1 (10^{11}$cm$^{-2})$.
The dashed lines are fits.  (b), (c) Slopes $\alpha$ and
amplitudes $\sigma(t=1$~s$)/\sigma_0$ of the relaxations,
respectively, \textit{vs.}
$\Delta n_s (10^{11}$cm$^{-2})$ for two fixed $n_0$ and
several $t_w$.  The values corresponding to $n_0 (10^{11}$cm$^{-2})=6.46$
were
multiplied by 10.  Open symbols:
device B. Dotted lines
guide the eye. (d) Sample B. $\sigma(t)$ for small $\Delta V_g$
scaled with $t_w$.
 \label{fig:DeltaV}}
\end{figure}

Full $t/t_w$ scaling is exhibited for all $n_s<n_c$.
However, as soon as $n_s\gtrsim n_c$, we find systematic
deviations from full aging [\textit{e.g.} Figs.~\ref{fig:mu}(a)
and \ref{fig:mu}(c)]. In
other glassy
materials~\cite{Struik,Ocio-mu}, it was found
%
%
that the data could be scaled with a modified waiting time
$(t_{w})^{\mu}$, where $\mu$ is a fitting parameter ($\mu=1$
for full aging).  Even though $\mu$ may not have a
clear physical meaning, the $\mu$-scaling approach has proved to
be useful for studying departures from full
aging~\cite{Struik,LesHouches-2002}.  By adopting a similar method,
we find that it is possible to achieve an approximate collapse of
the data [Figs.~\ref{fig:mu}(b) and \ref{fig:mu}(d)].
The plot of $\mu$ \textit{vs.} $n_0$ [Fig.~\ref{fig:mu}(f)] shows
a clear distinction between the full aging regime for $n_s<n_c$,
and the aging regime where significant departures from full
scaling are seen.  It is striking that the largest departure
occurs at $n_s\approx n_g$. For $n_s>n_g$,
it appears
as if the full aging is restored, but this may be an artifact of
trying to collapse very small ($\sigma/\sigma_{0}\lesssim 2$
percent) relaxations accompanied by instrumental (white) noise of
comparable magnitude [Fig.~\ref{fig:mu}(e)].  There are no
relaxations,
within the noise, for $V_0\geq 9$~V at $T=1$~K.
We have also determined that $\mu$ does not depend on $T$.

Another important issue that needs to be addressed is the role of
$V_1$ and of the step size $\Delta V_g=(V_1-V_0)$.  While
$\mu$ depends only on $V_0$ (provided $t_w\ll\tau_{eq}$ at $V_1$),
the amplitudes and the
slopes of the relaxations for a given $t_w$ do depend on
$\Delta V_g$.  Figure~\ref{fig:DeltaV}(a) shows such an example,
%
%
where $\sigma(t)$ are presented for fixed $V_0$, $t_w$, and
different $V_1$.  The data can be fitted by
$\sigma(t)/\sigma_0=[\sigma(t=1$~s$)/\sigma_0]\,t^{-\alpha}$ ($t$
in units of s) at least up to $t\sim t_w$.  The effect of
$\Delta V_g$ on the slope $\alpha$ and the amplitude
$\sigma(t=1$~s$)/\sigma_0$ is given in Figs.~\ref{fig:DeltaV}(b)
and \ref{fig:DeltaV}(c), respectively, for different $V_0$ and
$t_w$.  The increase of both quantities with
$\Delta V_g$ is not
too surprising, since larger $\Delta V_g$ lead to a higher number
of new electrons in the 2DES during $t_w$, taking the system
farther away from its original state at $V_0$.  It is intriguing,
though, that the results [Figs.~\ref{fig:DeltaV}(b) and
\ref{fig:DeltaV}(c)] seem to suggest that, for very small $\Delta
V_g$, both $\alpha$ and
$\log[\sigma(1$~s$)/\sigma_0]$ may become negative.  This
corresponds to $\sigma(t)$ approaching $\sigma_0$ from
``below'' instead of from above as in
Fig.~\ref{fig:full}(c)~\cite{comment-noOS}. Indeed, we have
observed instances of such behavior, as shown in
Fig.~\ref{fig:DeltaV}(d) for the case where both $V_0$ and $V_1$
happen to be in the insulating regime.  The nonmonotonic dependence of the
system response, here characterized by $\alpha$ and
$\sigma(1$~s$)/\sigma_0$, on the perturbation $\Delta V_g$,
bears a remarkable resemblance to the results of negative $T$ cycle spin
glass experiments (see Figs. 1--3 in Ref.~\cite{over-spin}), where
the so-called ``memory anomaly'' at low $\Delta T$ was
interpreted~\cite{over-spin} based on the hierarchical
distribution of states in the free energy landscape.  The noise
studies in the aging regime~\cite{JJPRL} have already provided
support for the hierarchical picture of glassiness in the
2DES.  Unfortunately, it is not possible to perform systematic
studies of aging for small, including negative, $\Delta V_g$ in a
2DES, because then $\sigma(1$~s$)/\sigma_0$ is typically
comparable to the large noise that is intrinsic to aging.

\begin{figure}
\centerline{\epsfig{file=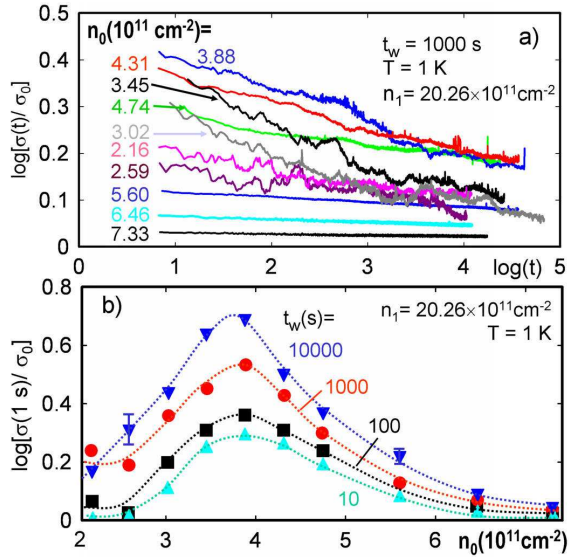,width=7.5cm,clip=}}
\caption{(color online) (a) $\sigma(t)$ for fixed $t_w$, $n_1$,
and several $n_0$.  (b) Relaxation amplitudes
\textit{vs.} $n_0$ for several $t_w$.  Dotted lines guide the eye.
The error bars typical of low and high $n_0$ values are also shown.
 \label{fig:amp}}
\end{figure}

The raw data [\textit{e.g.} Figs.~\ref{fig:mu}(a),
\ref{fig:mu}(c), and \ref{fig:mu}(e)], as well as
Figs.~\ref{fig:DeltaV}(b) and \ref{fig:DeltaV}(c), indicate also
large variations in $\sigma(1$~s$)/\sigma_0$ and $\alpha$ as a
function of $V_0$ for fixed $t_w$ and $\Delta V_g$.  This was
explored in
detail by varying $V_0$, while keeping $t_w$
and $V_1$ fixed [Fig.~\ref{fig:amp}(a)].
We expect, based on
%
%
Fig.~\ref{fig:DeltaV}, to see an increase of both quantities with
decreasing $V_0$, since $\Delta V_g$ is getting larger.  While an
increase is observed [Fig.~\ref{fig:amp}(b)], it is much stronger
than expected.  For example,  even for the same
$\Delta n_s(10^{11}$cm$^{-2})=13.79$ and $t_w=1000$~s,
$\log[\sigma(n_0(10^{11}$cm$^{-2})=3.88$, $t=$1~s$)/\sigma_0]\approx 0.4$
[Fig.~\ref{fig:DeltaV}(c)], \textit{i.e.} still an
order of magnitude larger than
for
$n_0 (10^{11}$cm$^{-2})=6.46$.
Even more striking, however,
is the dramatic
\textit{decrease} of $\sigma(1$~s$)/\sigma_0$
[Fig.~\ref{fig:amp}(b)] as $n_0$ is reduced further, contrary to
simple expectations based on Fig.~\ref{fig:DeltaV}.
The slopes $\alpha$ exhibit the same nonmonotonic behavior [Fig.~\ref{fig:amp}(a)].  Again, the abrupt change in the aging
properties occurs at $n_s\approx n_c$: the relaxation amplitude peaks just before the system becomes metallic, and it
is suppressed upon going deeper into the insulator.  This
is
reminiscent of the suppression of the fluctuations of
$\sigma(t)$ observed earlier in the same system at $n_c$ (Ref.~\cite{SBPRL}, Fig.~4
inset).  While a clear understanding of these
two effects is lacking, it is plausible that collective
charge rearrangements that are responsible for the slow dynamics
will be suppressed as the 2DES becomes strongly localized.  It is
interesting to speculate whether this is related in any way to the
problem of many-body localization~\cite{Basko}.
 It would be also of interest to study aging deeper in the insulator,
in the variable-range hopping regime, but that
is not possible because of the small relaxation amplitudes and the
large intrinsic sample noise.  The effects of disorder could be explored further by extending the relaxation studies to cleaner 2DES, where $n_g\gtrsim n_c$~\cite{JJPRL}.

In summary, a detailed study of aging in a 2DES in Si
shows an abrupt change in the nature of the glassy phase at
the 2D MIT before it vanishes entirely at a higher density
$n_g$.  The results put constraints on the theories of
glassy freezing and its role in the physics of the 2D MIT.

We are grateful to I. Rai\v{c}evi\'{c} for technical
help, V. Dobrosavljevi\'c for discussions, NSF DMR-0403491 and
NHMFL via NSF DMR-0084173 for financial support.

\newcommand{\noopsort}[1]{} \newcommand{\printfirst}[2]{#1}
  \newcommand{\singleletter}[1]{#1} \newcommand{\switchargs}[2]{#2#1}


\begin{thebibliography}{24}
\bibitem[{kfl()}]{kfl}
\bibinfo{note}{Relaxations at finite $T$ ($T>T_g=0$) are visible as long as $k_{F}l<1$ ($k_{F}$ -- Fermi wave vector, $l$ -- mean free
path)}.
\bibitem{tw-comment} This expression does not hold in the $t_{w}\rightarrow 0$ limit, since the power-law $\sigma(t)$ does not hold for $t/t_{w}>0$.
\bibitem{comment-noOS}
This is unrelated to
``overshooting'' when 
$\tau_{eq}\ll~\!t_w$
~\cite{tw-PRL}.
\end{thebibliography}
\end{document}